\documentclass[12pt]{iopart}

\usepackage{graphicx}
\begin{document}

\title[Event-by-event mean $p_{\rm T}$ fluctuations measured by the ALICE experiment]{\large{Event-by-event mean $p_{\rm T}$ fluctuations in pp and Pb--Pb collisions measured by the ALICE experiment at the LHC}}
\author{Stefan Heckel$^1$ for the ALICE collaboration}
\address{$^1$ Institut f\"{u}r Kernphysik, Goethe-Universit\"{a}t, Max-von-Laue-Str. 1, 60438 Frankfurt am Main, Germany}
\ead{sheckel@ikf.uni-frankfurt.de}

\begin{abstract}
Non-statistical event-by-event fluctuations of the mean transverse momentum of charged particles in pp and Pb--Pb collisions are studied using the ALICE experiment at the LHC.
Little collision energy dependence is observed in pp. The data indicate a common scaling behaviour with event multiplicity from pp to semi-central Pb--Pb collisions. In central Pb--Pb, the results deviate from this trend, exhibiting a significant reduction of the fluctuation strength. The results are compared with measurements in Au--Au collisions at lower energies and with Monte Carlo simulations.
\end{abstract}

\section{Introduction}

Event-by-event fluctuations of mean transverse momentum $\langle p_{\rm T} \rangle$ contain information on the dynamics and correlations in pp and heavy-ion collisions. Measurements in pp serve as a baseline representing ``known'' physics like $p_{\rm T}$ correlations due to resonance decays, the Hanbury Brown Twiss (HBT) effect or (mini-)jets. In heavy-ion collisions, fluctuations may also be related to critical behaviour of the system in the vicinity of a phase boundary \cite{stephanov1,stephanov2} or the occurrence of thermalization and collectivity \cite{gavin}. In general, such phenomena in AA can be observed by a variation (increase or reduction) of the fluctuation pattern with respect to the pp baseline.

In this analysis, non-statistical event-by-event fluctuations of the mean transverse momentum of charged particles in pp and Pb--Pb collisions are studied using the ALICE detector at the LHC \cite{detector}. The analysis is performed in $0.15 < p_{\rm T} < 2$~GeV/c and $|\eta| < 0.8$. Multiplicity dependent results are obtained for pp collisions at $\sqrt{s}$~=~0.9, 2.76 and 7~TeV. Pb--Pb collisions at $\sqrt{s_{\rm NN}}$~=~2.76~TeV are analysed in intervals of multiplicity and centrality, the latter in 5\% intervals of the total hadronic Pb--Pb cross section \cite{toia}.

\begin{figure}[t]
	\centering
		\includegraphics[width=0.55\textwidth]{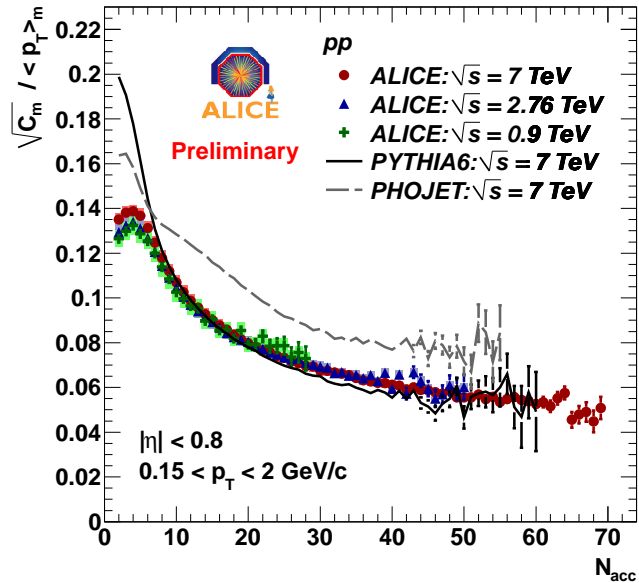}
	\caption{Relative fluctuations $\sqrt{C_m} / \langle p_{\rm T} \rangle_m$ as a function of accepted multiplicity $N_{\rm acc}$ in pp collisions at different LHC energies and comparison to event generators at $\sqrt{s}$~=~7~TeV: PYTHIA6 with Perugia0 tune (solid line) and PHOJET (dashed line).}
	\label{fig:PtFluc_figure1}
\end{figure}

\section{Analysis}

For charged particle track and vertex reconstruction the Time Projection Chamber (TPC) is used \cite{tpc}. The Inner Tracking System (ITS) serves to improve the vertex reconstruction in pp collisions. Inelastic pp and Pb--Pb events are selected with a minimum bias interaction trigger \cite{toia}.

Event-by-event fluctuations in heavy-ion collisions are composed of statistical and dynamical contributions, the latter consisting of effects seen already in pp collisions, together with effects that are specific to AA, i.e. $\sigma_{\rm total}^2 = \sigma_{\rm stat}^2 + \sigma_{\rm dyn}^2$.

The two-particle correlator $C_m = \langle \Delta p_{{\rm T},i}, \Delta p_{{\rm T},j} \rangle$ is a measure of the dynamical component $\sigma_{\rm dyn}^2$ and therefore is well suited for an event-by-event analysis of $\langle p_{\rm T} \rangle$ fluctuations \cite{voloshin,star}. It is the mean of covariances of all pairs of particles $i$ and $j$ in the same event with respect to the inclusive $\langle p_{\rm T} \rangle$ in a given multiplicity class $m$ and is defined as
\begin{equation}
	C_m = \frac{1}{\sum_{k=1}^{n_{\rm ev}}{N_{k}^{\rm pairs}}} \cdot \sum_{k=1}^{n_{\rm ev}} \sum_{i=1}^{N_{k}} \sum_{j=i+1}^{N_{k}} (p_{{\rm T},i} - \langle p_{\rm T} \rangle_m ) \cdot (p_{{\rm T},j} - \langle p_{\rm T} \rangle_m ),
	\label{eq:correlator}
\end{equation}

where $n_{\rm ev}$ is the number of events in a given multiplicity class $m$, $N_{k}$ is the number of particles, $N_{k}^{\rm pairs} = 0.5 \cdot N_{k} \cdot (N_{k}-1)$ the number of particle pairs in event $k$ and $\langle p_{\rm T} \rangle_m$ the average $p_{\rm T}$ of all tracks of all events in class $m$. By construction, $C_m = 0$ in the presence of only statistical fluctuations. The correlator is calculated in bins of the event multiplicity $N_{\rm acc}$, where $N_{\rm acc}$ is the number of tracks accepted by the analysis cuts.

\begin{figure}[t]
	\centering
		\includegraphics[width=0.55\textwidth]{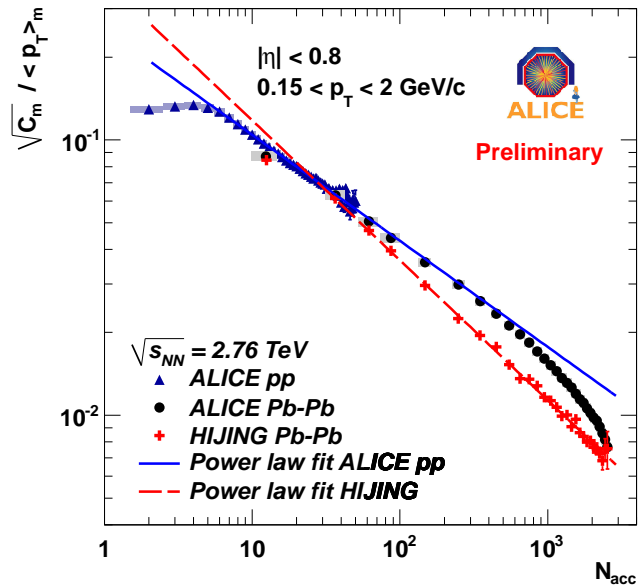}
	\caption{Relative fluctuations $\sqrt{C_m} / \langle p_{\rm T} \rangle_m$ as a function of accepted multiplicity $N_{\rm acc}$ in pp and Pb--Pb collisions at $\sqrt{s_{\rm NN}}$~=~2.76~TeV and comparison to the HIJING event generator. The solid line represents a power law fit through the pp data points in the interval $8 \leq N_{\rm acc} \leq 40$ while the dashed line is a power law fit through the HIJING points in the interval $25 \leq N_{\rm acc} \leq 2600$.}
	\label{fig:PtFluc_figure2}
\end{figure}

\section{Results}

In the following, the results are presented in terms of the relative fluctuations $\sqrt{C_m} / \langle p_{\rm T} \rangle_m$. Figure~\ref{fig:PtFluc_figure1} shows the results for pp collisions at $\sqrt{s}$~=~0.9, 2.76 and 7~TeV together with Monte Carlo simulations using PYTHIA6 \cite{pythia} (Perugia0 tune) and PHOJET \cite{phojet} at $\sqrt{s}$~=~7~TeV. Significant non-statistical fluctuations are observed, demonstrating a dilution with increasing multiplicity. In this relative representation the data look universal at LHC energies, except at small multiplicities, where the fluctuations increase slightly with beam energy. PYTHIA6 gives a reasonable description of the data for $N_{\rm acc} \geq 7$, though showing a slightly stronger decrease with multiplicity. PHOJET does not describe the data well.

Pb--Pb collisions also show significant non-statistical fluctuations and dilution with increasing multiplicity as is presented in figure~\ref{fig:PtFluc_figure2} together with pp data and a comparison to the HIJING \cite{hijing} event generator, all at $\sqrt{s_{\rm NN}}$~=~2.76~TeV. The pp data are fitted with a power law $\sqrt{C_m} / \langle p_{\rm T} \rangle_m (N_{\rm acc}) = A \cdot N_{\rm acc}^{b}$, with $b = -0.385\pm0.003$. The Pb--Pb data agree also well with this pp baseline up to $N_{\rm acc} \approx 600$, justifying the use of a common parametrization. Central Pb--Pb collisions show a significant additional reduction of the fluctuations. HIJING does not describe the shape of the Pb--Pb data. A power law fit to the HIJING data yields $b = -0.508\pm0.004$, i.e. consistent with scaling by $1 / \sqrt{N_{\rm acc}}$. The HIJING calculation shows no deviation from the power law over the full multiplicity range. Further studies are needed to identify possible mechanisms leading to the non-trivial reduction of $\langle p_{\rm T} \rangle$ fluctuations observed in central Pb--Pb collisions.

\begin{figure}[t]
	\centering
		\includegraphics[width=0.90\textwidth]{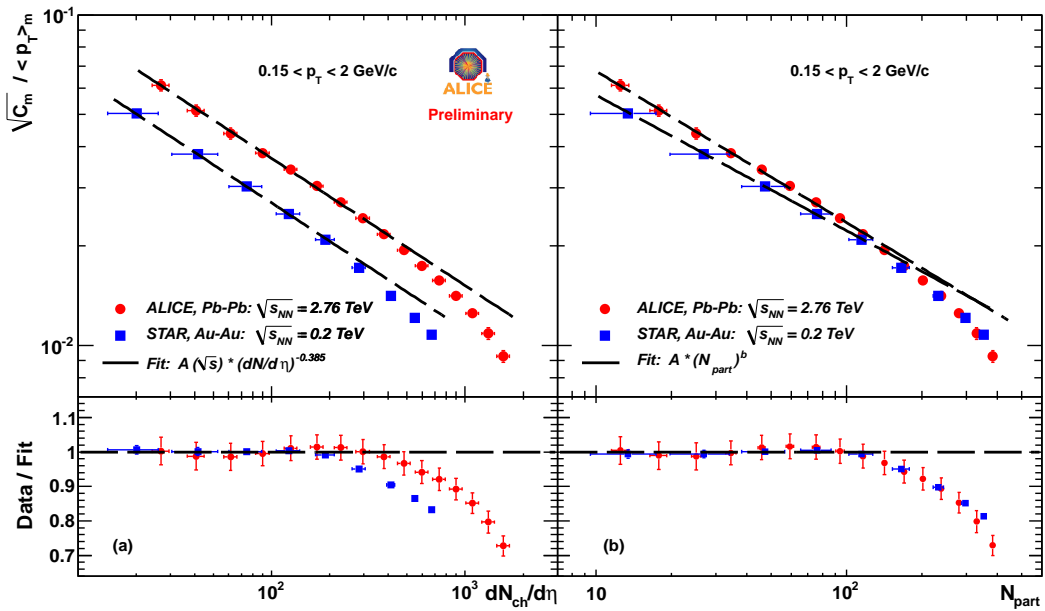}
	\caption{Relative fluctuations in Pb--Pb collisions measured by ALICE and Au-Au collisions measured by STAR as a function of $\rmd N_{\rm ch} / \rmd \eta$ (a) and $N_{\rm part}$ (b). The lower panels show comparisons to power law fits (see text). The horizontal error bars indicate systematic uncertainties on $\rmd N_{\rm ch} / \rmd \eta$ and $N_{\rm part}$. Vertical error bars on the ALICE data include statistical and systematic errors, added in quadrature. The error bars on the STAR data are from \cite{star}.}
	\label{fig:PtFluc_figure3}
\end{figure}

A comparison to STAR Au--Au data at $\sqrt{s_{\rm NN}}$~=~0.2~TeV \cite{star} is shown in figure~\ref{fig:PtFluc_figure3} (a) in terms of the charged particle multiplicity density $\rmd N_{\rm ch} / \rmd \eta$ \cite{dndeta} and (b) in terms of the number of participants $N_{\rm part}$. In (a) both data sets are fitted by a power law, fixing the power to $b = -0.385$ as obtained from the ALICE pp data at $\sqrt{s}$~=~2.76~TeV. Both data sets agree well with this parametrization in peripheral events, suggesting the applicability of our pp baseline also for RHIC AA data. At large multiplicities, the STAR data deviate also from the baseline, but the deviation is more pronounced in the ALICE data. Figure~\ref{fig:PtFluc_figure3} (b) shows the ALICE and STAR data versus $N_{\rm part}$, together with free power law fits to the peripheral data. The data sets cannot be described by a common exponent; however, the deviation from the fit in central events is very similar, pointing to a possible scaling behaviour of $\langle p_{\rm T} \rangle$ fluctuations in terms of $N_{\rm part}$.

\end{document}